\documentclass[aps,prl,twocolumn,showpacs,amsmath,amssymb,bm,superscriptaddress]{revtex4-1}  
\usepackage{graphicx}  
\usepackage{dcolumn}   
\usepackage{hyperref}
\usepackage{bm}        

\newcommand{\Dsm}{\ensuremath{D_s^-}}
\newcommand{\Dsp}{\ensuremath{D_s^+}}
\newcommand{\Ds}{\ensuremath{D_s^\pm}}

\newcommand{\Dsdecay}{\ensuremath{D_s^\pm \rightarrow \phi \pi^{\pm}}}
\newcommand{\DsPdecay}{\ensuremath{D_s^+ \rightarrow \phi \pi^{+}}}
\newcommand{\DsMdecay}{\ensuremath{D_s^- \rightarrow \phi \pi^{-}}}
\newcommand{\Acp}{\ensuremath{A_{CP}}}
\newcommand{\fb}{\ensuremath{\rm{fb}^{-1}}}

\newcommand {\asls} {\ensuremath{a^s_{\mathrm{sl}}}}
\newcommand {\delg} {\ensuremath{\Delta\Gamma_d/\Gamma_d}}

\newcommand {\ks} {\ensuremath{K^0_S}}

\newcommand {\Bd} {\ensuremath{B^0_d}}
\newcommand {\Bs} {\ensuremath{B^0_s}}

\newcommand {\barBs} {\ensuremath{\bar{B}^0_s}}

\begin{document}

\hspace{5.2in} \mbox{FERMILAB-PUB-13-550-E}

\title{
Measurement of the direct \textit{CP}-violating charge asymmetry in $\bm{\Ds \rightarrow \phi \pi^{\pm}}$ decays}

\affiliation{LAFEX, Centro Brasileiro de Pesquisas F\'{i}sicas, Rio de Janeiro, Brazil}
\affiliation{Universidade do Estado do Rio de Janeiro, Rio de Janeiro, Brazil}
\affiliation{Universidade Federal do ABC, Santo Andr\'e, Brazil}
\affiliation{University of Science and Technology of China, Hefei, People's Republic of China}
\affiliation{Universidad de los Andes, Bogot\'a, Colombia}
\affiliation{Charles University, Faculty of Mathematics and Physics, Center for Particle Physics, Prague, Czech Republic}
\affiliation{Czech Technical University in Prague, Prague, Czech Republic}
\affiliation{Institute of Physics, Academy of Sciences of the Czech Republic, Prague, Czech Republic}
\affiliation{Universidad San Francisco de Quito, Quito, Ecuador}
\affiliation{LPC, Universit\'e Blaise Pascal, CNRS/IN2P3, Clermont, France}
\affiliation{LPSC, Universit\'e Joseph Fourier Grenoble 1, CNRS/IN2P3, Institut National Polytechnique de Grenoble, Grenoble, France}
\affiliation{CPPM, Aix-Marseille Universit\'e, CNRS/IN2P3, Marseille, France}
\affiliation{LAL, Universit\'e Paris-Sud, CNRS/IN2P3, Orsay, France}
\affiliation{LPNHE, Universit\'es Paris VI and VII, CNRS/IN2P3, Paris, France}
\affiliation{CEA, Irfu, SPP, Saclay, France}
\affiliation{IPHC, Universit\'e de Strasbourg, CNRS/IN2P3, Strasbourg, France}
\affiliation{IPNL, Universit\'e Lyon 1, CNRS/IN2P3, Villeurbanne, France and Universit\'e de Lyon, Lyon, France}
\affiliation{III. Physikalisches Institut A, RWTH Aachen University, Aachen, Germany}
\affiliation{Physikalisches Institut, Universit\"at Freiburg, Freiburg, Germany}
\affiliation{II. Physikalisches Institut, Georg-August-Universit\"at G\"ottingen, G\"ottingen, Germany}
\affiliation{Institut f\"ur Physik, Universit\"at Mainz, Mainz, Germany}
\affiliation{Ludwig-Maximilians-Universit\"at M\"unchen, M\"unchen, Germany}
\affiliation{Panjab University, Chandigarh, India}
\affiliation{Delhi University, Delhi, India}
\affiliation{Tata Institute of Fundamental Research, Mumbai, India}
\affiliation{University College Dublin, Dublin, Ireland}
\affiliation{Korea Detector Laboratory, Korea University, Seoul, Korea}
\affiliation{CINVESTAV, Mexico City, Mexico}
\affiliation{Nikhef, Science Park, Amsterdam, the Netherlands}
\affiliation{Radboud University Nijmegen, Nijmegen, the Netherlands}
\affiliation{Joint Institute for Nuclear Research, Dubna, Russia}
\affiliation{Institute for Theoretical and Experimental Physics, Moscow, Russia}
\affiliation{Moscow State University, Moscow, Russia}
\affiliation{Institute for High Energy Physics, Protvino, Russia}
\affiliation{Petersburg Nuclear Physics Institute, St. Petersburg, Russia}
\affiliation{Instituci\'{o} Catalana de Recerca i Estudis Avan\c{c}ats (ICREA) and Institut de F\'{i}sica d'Altes Energies (IFAE), Barcelona, Spain}
\affiliation{Uppsala University, Uppsala, Sweden}
\affiliation{Lancaster University, Lancaster LA1 4YB, United Kingdom}
\affiliation{Imperial College London, London SW7 2AZ, United Kingdom}
\affiliation{The University of Manchester, Manchester M13 9PL, United Kingdom}
\affiliation{University of Arizona, Tucson, Arizona 85721, USA}
\affiliation{University of California Riverside, Riverside, California 92521, USA}
\affiliation{Florida State University, Tallahassee, Florida 32306, USA}
\affiliation{Fermi National Accelerator Laboratory, Batavia, Illinois 60510, USA}
\affiliation{University of Illinois at Chicago, Chicago, Illinois 60607, USA}
\affiliation{Northern Illinois University, DeKalb, Illinois 60115, USA}
\affiliation{Northwestern University, Evanston, Illinois 60208, USA}
\affiliation{Indiana University, Bloomington, Indiana 47405, USA}
\affiliation{Purdue University Calumet, Hammond, Indiana 46323, USA}
\affiliation{University of Notre Dame, Notre Dame, Indiana 46556, USA}
\affiliation{Iowa State University, Ames, Iowa 50011, USA}
\affiliation{University of Kansas, Lawrence, Kansas 66045, USA}
\affiliation{Louisiana Tech University, Ruston, Louisiana 71272, USA}
\affiliation{Northeastern University, Boston, Massachusetts 02115, USA}
\affiliation{University of Michigan, Ann Arbor, Michigan 48109, USA}
\affiliation{Michigan State University, East Lansing, Michigan 48824, USA}
\affiliation{University of Mississippi, University, Mississippi 38677, USA}
\affiliation{University of Nebraska, Lincoln, Nebraska 68588, USA}
\affiliation{Rutgers University, Piscataway, New Jersey 08855, USA}
\affiliation{Princeton University, Princeton, New Jersey 08544, USA}
\affiliation{State University of New York, Buffalo, New York 14260, USA}
\affiliation{University of Rochester, Rochester, New York 14627, USA}
\affiliation{State University of New York, Stony Brook, New York 11794, USA}
\affiliation{Brookhaven National Laboratory, Upton, New York 11973, USA}
\affiliation{Langston University, Langston, Oklahoma 73050, USA}
\affiliation{University of Oklahoma, Norman, Oklahoma 73019, USA}
\affiliation{Oklahoma State University, Stillwater, Oklahoma 74078, USA}
\affiliation{Brown University, Providence, Rhode Island 02912, USA}
\affiliation{University of Texas, Arlington, Texas 76019, USA}
\affiliation{Southern Methodist University, Dallas, Texas 75275, USA}
\affiliation{Rice University, Houston, Texas 77005, USA}
\affiliation{University of Virginia, Charlottesville, Virginia 22904, USA}
\affiliation{University of Washington, Seattle, Washington 98195, USA}
\author{V.M.~Abazov} \affiliation{Joint Institute for Nuclear Research, Dubna, Russia}
\author{B.~Abbott} \affiliation{University of Oklahoma, Norman, Oklahoma 73019, USA}
\author{B.S.~Acharya} \affiliation{Tata Institute of Fundamental Research, Mumbai, India}
\author{M.~Adams} \affiliation{University of Illinois at Chicago, Chicago, Illinois 60607, USA}
\author{T.~Adams} \affiliation{Florida State University, Tallahassee, Florida 32306, USA}
\author{J.P.~Agnew} \affiliation{The University of Manchester, Manchester M13 9PL, United Kingdom}
\author{G.D.~Alexeev} \affiliation{Joint Institute for Nuclear Research, Dubna, Russia}
\author{G.~Alkhazov} \affiliation{Petersburg Nuclear Physics Institute, St. Petersburg, Russia}
\author{A.~Alton$^{a}$} \affiliation{University of Michigan, Ann Arbor, Michigan 48109, USA}
\author{A.~Askew} \affiliation{Florida State University, Tallahassee, Florida 32306, USA}
% ...
%
%
\collaboration{The D0 Collaboration\footnote{with visitors from
$^{a}$Augustana College, Sioux Falls, SD, USA,
$^{b}$The University of Liverpool, Liverpool, UK,
$^{c}$DESY, Hamburg, Germany,
$^{d}$Universidad Michoacana de San Nicolas de Hidalgo, Morelia, Mexico
$^{e}$SLAC, Menlo Park, CA, USA,
$^{f}$University College London, London, UK,
$^{g}$Centro de Investigacion en Computacion - IPN, Mexico City, Mexico,
$^{h}$Universidade Estadual Paulista, S\~ao Paulo, Brazil,
$^{i}$Karlsruher Institut f\"ur Technologie (KIT) - Steinbuch Centre for Computing (SCC)
and
$^{j}$Office of Science, U.S. Department of Energy, Washington, D.C. 20585, USA.
}} \noaffiliation
\vskip 0.25cm

\date{December 3, 2013}

\begin{abstract}
We present a  measurement of the direct \textit{CP}-violating charge
asymmetry in $\Ds \to \phi \pi^{\pm}$  decays where the $\phi$ meson
decays to $K^+K^-$, using the full Run II data set with an integrated
luminosity  of 10.4 fb$^{-1}$ of proton-antiproton collisions collected
using the D0 detector at the Fermilab Tevatron Collider. 
The normalized difference \Acp\ in the yield of \Dsp\ and \Dsm\ mesons
in these decays is measured by fitting the difference between their 
reconstructed invariant mass distributions. This results in an asymmetry 
of $\Acp =\left[ -0.38 \pm 0.27 \right]\%$, which is the most precise 
measurement of this quantity to date. The result is consistent with the 
standard model prediction of zero \textit{CP} asymmetry in this decay.
\end{abstract}

\maketitle

Direct {\it CP} violation (CPV) in the Cabbibo-preferred charm decay
\Dsdecay\ should be non-existent in  the standard model (SM). In the SM,
direct CPV will occur if there are tree and loop (penguin) processes
that can interfere with different strong and weak phases. There will be
no CPV in \Dsdecay\ decays as all of the contributing processes have the
same weak phase ($V_{cs}V_{ud}$)~\cite{CPviolation}. Any  CPV in this
channel would indicate the existence of physics beyond the SM 
(for examples, see Ref.~\cite{lhcbPhysics}). The most
recent investigation of this decay by the CLEO Collaboration yields a
{\it CP}-violating charge asymmetry of 
$A_{CP}\left(\Dsdecay \right) = 
\left[ -0.5 \pm 0.8 \thinspace (\rm{stat}) \pm 0.4 \thinspace (\rm{syst}) \right]\%$~\cite{cleo2013} where
the direct CPV charge asymmetry in the decay \Dsdecay\ is defined as
\begin{equation}
\Acp\  = 
\frac{\Gamma\left(\DsPdecay \right) - \Gamma\left(\DsMdecay \right)}
     {\Gamma\left(\DsPdecay \right) + \Gamma\left(\DsMdecay \right)}. 
\end{equation}

No CPV in this decay is assumed in measurements of the
time-integrated flavor-specific semileptonic charge asymmetry in the
decays  of oscillating  neutral $B_s^0$ mesons using the decay $(\barBs) 
\to \Bs \to D_{s} \mu X$
by the D0~\cite{d0assl} and LHCb~\cite{lhcbassl} Collaborations, and 
in the search for direct CPV in
$D^+ \to \phi \pi^+$ and $D_s^+ \to K_S^0 \pi^+$ decays by the LHCb
Collaboration~\cite{lhcb2013}. Assuming no CPV in \Dsdecay\ decays, 
the LHCb Collaboration finds that the production asymmetry
of \Dsdecay\ decays in proton-proton
interactions is $A_{\rm prod} = 
\left(\sigma(\Dsp) - \sigma(\Dsm)\right)/\left(\sigma(\Dsp) + \sigma(\Dsm)\right)
=  \left[ -0.33 \pm 0.22 \thinspace (\rm{stat}) \pm
0.10 \thinspace (\rm{syst}) \right]\%$~\cite{lhcbprod} 
where $\sigma(\Ds)$ is the inclusive prompt production cross-section. 
D0 is the only experiment which can test this assumption with
sufficient sensitivity in the foreseeable future since the 
Tevatron collides protons on anti-protons  which is a 
\textit{CP}-invariant initial state,
and that the systematic uncertainties for this process are small at D0  due to the 
specific features of the detector.

A measure of the CPV in mixing is obtained from the 
average of the direct measurements of the semileptonic charge
asymmetry in decays of neutral $B_s^0$ mesons using the decay $(\barBs) 
\to \Bs \to D_{s} \mu X$~\cite{d0assl,lhcbassl} yielding  $\asls =
\left[ -0.50 \pm 0.52 \right]\%$. 
This asymmetry can also be extracted
indirectly from measurements of charge asymmetries of single muons and 
like-sign dimuons~\cite{dimuon2013}, the semileptonic charge asymmetry 
of neutral $B_d^0$ mesons (using the average at the
$\Upsilon(4S)$~\cite{hfag} and the D0 result~\cite{d0adsl}), and the
ratio of the decay width difference and the average decay width of
$B_d^0$, \delg~\cite{hfag} resulting in  $\asls (\rm{indirect})= \left[
-1.46 \pm 0.78 \right]\%$. 
    While the difference between these two asymmetries is not significant, 
    CPV in \Dsdecay\ decays could potentially explain the 3.6 standard deviation 
    discrepancy between the SM prediction and the measured charge asymmetries of 
    like-sign dimuons~\cite{dimuon2013} given that no such discrepancy has been observed in 
    direct measurements of semileptonic charge asymmetries.

In this Letter, the D0 Collaboration presents a measurement of \Acp\
using the full Tevatron Run II data sample with an integrated luminosity
of 10.4 \fb . We assume there is negligible net production asymmetry between
$D_s^+$ and $D_s^-$ mesons in proton-antiproton collisions.  We also
assume that any integrated production asymmetry of $b$ hadrons that 
decay to \Ds\ is negligible.

This measurement of \Acp\ makes use of the methods for extracting
asymmetries used in the D0 analyses of the time-integrated flavor-specific
semileptonic charge asymmetry in the decays of neutral $B$
mesons~\cite{d0assl,d0adsl}. We measure the raw asymmetry
\begin{equation}
 A_{D_s}  = \frac{N_{D_s^+} - N_{D_s^-}  }{N_{D_s^+} + N_{D_s^-} },
\label{raw}
\end{equation}
\noindent where $N_{D_s^+}$ ($N_{D_s^-}$) is the number of reconstructed
\DsPdecay\ (\DsMdecay ) decays. The charge asymmetry in $D_s^\pm$ decays
is then given by (neglecting any terms second- or higher-order in the
asymmetry) 
\begin{equation}
\label{eq:asymm}
\Acp = A_{D_s} - A_{\rm det} - A_{\rm phys},
\end{equation}
\noindent where $A_{\rm det}$ is due to residual reconstruction
asymmetries in the detector, and $A_{\rm phys}$ is the  charge asymmetry 
resulting from the decay of $b$ hadrons  to \Ds\ mesons.

The D0 detector has a central tracking system consisting of a silicon
microstrip tracker and the central fiber tracker, both located within a
2~T superconducting solenoidal magnet~\cite{d0det, layer0}. A muon
system, covering $|\eta|<2$~\cite{eta}, consists of a layer of tracking
detectors and scintillation trigger counters in front of 1.8~T toroidal
magnets, followed by two similar layers after the
toroids~\cite{run2muon}. 

The polarities of the toroidal and solenoidal magnetic fields are
reversed on average every two weeks so that the four solenoid-toroid
polarity combinations are exposed to approximately the same integrated
luminosity. This allows for a cancellation of first-order effects related
to instrumental charge and momentum reconstruction 
asymmetries. To ensure a more complete cancellation of the
uncertainties, the events are weighted according to the number of $\phi
\pi^\pm$ decays collected in each configuration of the magnets' polarities 
(polarity-weighting). The weighting is based on the number of events 
containing \Ds\ decay products that pass the selection
criteria and the likelihood selection (described below), and that are in
the $\phi \pi^\pm$ invariant mass range used for the fit.

As there was no dedicated trigger for hadronic decays of heavy flavor mesons,
the data were collected with a suite of single and dimuon triggers. The
trigger and offline streaming requirements bias the composition of the 
data. The muon requirement will preferentially select events 
with semileptonic decays and may enhance the contribution of events 
produced by the decay of $b$ hadrons.
The effect of this bias is corrected using a Monte Carlo 
(MC) simulation (described below).

The $D_s^\pm \rightarrow  \phi \pi^\pm$; $\phi \rightarrow K^+ K^-$
decay is reconstructed as follows. Since the D0 detector is unable 
to distinguish between charged $K$ and $\pi$ mesons, the two particles from the $\phi$
decay are assumed to be kaons and are required to have  $p_T >
0.7$~GeV/$c$, opposite charge and a reconstructed invariant mass of
$M(K^+K^-) < 1.07$~GeV/$c^2$. The third particle, assumed to be the
charged pion, is required to have  $p_T > 0.5$~GeV/$c$. The three
particles are combined to create a common $D_s^\pm$ decay vertex using
the algorithm described in Ref.~\cite{vertex}. The  cosine of the angle
between the $D^\pm_s$ momentum and the vector from the $p\bar{p}$
collisions vertex to the $D^\pm_s$ decay vertex in the transverse plane
is required to be greater than 0.95. The trajectories of the $D^\pm_s$
candidate tracks are required to be consistent with originating from a
common vertex  and to have an invariant mass of  $1.7 < M(K^+K^-\pi^\pm)
< 2.3$~GeV$/c^2$.  To reduce combinatorial background, the $D^\pm_s$
vertex is required to have a displacement from the $p\bar{p}$ collision
vertex  in the transverse plane with a significance of at least four
standard deviations.

To improve the significance of the $D_s^\pm$ selection, we use a
likelihood ratio~\cite{like_ratio} to combine several variables that 
discriminate between signal and the combinatoric background: the helicity angle between
the $D_s^\pm$ and $K^{\mp}$ momenta in the center-of-mass frame of the
$\phi$ meson; the isolation of the $D_s^\pm$ system, defined as
$I=\left|\vec{p}(D_s^\pm)\right|/[\left|\vec{p}(D_s^\pm)\right|+\Sigma \left|\vec{p}_i \right|]$, 
where $\vec{p}(D_s^\pm)$ is the vector sum of the momenta of the three tracks that
make up the $D_s^\pm$ meson and $\Sigma \left|\vec{p}_i\right|$  is the sum of
momenta of all charged particles not associated with the $D_s^\pm$ meson
in a cone of $\sqrt{(\Delta \phi)^2+(\Delta \eta)^2}<0.5$ around the
$D_s^\pm$ direction~\cite{eta}; the $\chi^2$ of the $D_s^\pm$ vertex
fit; the invariant mass $M(K^+ K^-)$; $p_T(K^+ K^-)$; the cosine of the
angle between the $D^\pm_s$ momentum and the vector from the $p\bar{p}$
collision vertex to the $D^\pm_s$ decay vertex, and the separation
between the $K^\pm$ and $\pi^\pm$ mesons with the same charge, defined
as $\sqrt{(\phi_K - \phi_\pi)^2 + (\eta_K - \eta_\pi)^2}$. 
The signal is modelled using a MC simulation of \Dsdecay\ decays and 
the background is modelled using the data  (which is dominated by 
background events) before applying the likelihood selection.  
The requirement on the likelihood ratio variable is chosen to minimize the
statistical uncertainty on \Acp\ obtained using the signal extraction
procedure described below.

\begin{figure}[htbp]
\includegraphics[width=1.0\columnwidth]{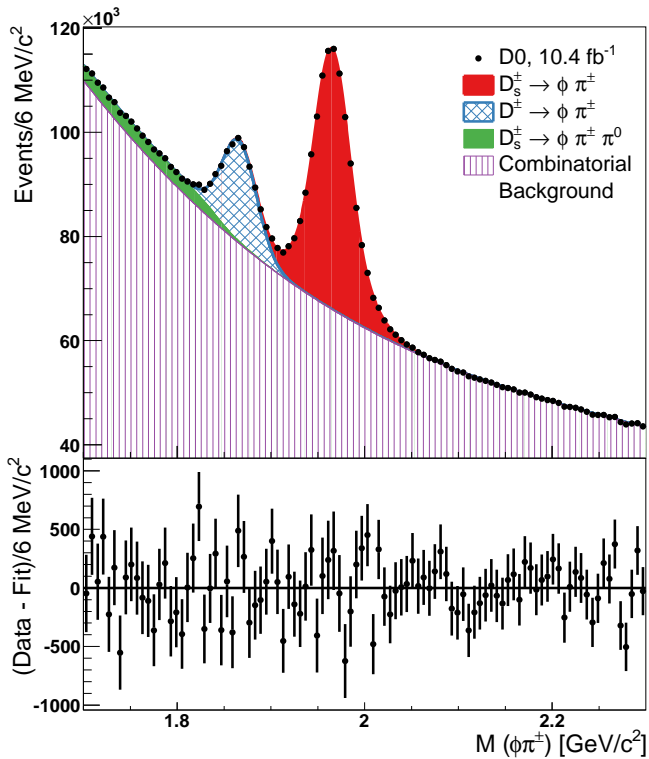}
\caption{\label{Fig:WeightedDSCandidates} 
The polarity-weighted $\phi \pi^\pm$ invariant mass distribution.
 The lower mass peak is due to  the decay $D^{\pm} \rightarrow \phi
 \pi^\pm$ while the second peak is due to the $D_s^{\pm}$ meson decay.
 Note the zero-suppression on the vertical axis. The bottom panel shows
 the fit residuals. The error bars represent the statistical
 uncertainties. 
}
\end{figure}

The $M(K^+K^-\pi^\pm)$  distribution is displayed in bins of
6~MeV$/c^2$ over a  range of $1.7 <  M(K^+K^-\pi^\pm) < 2.3$~GeV$/c^2$,
and the number of signal and background events is extracted by a $\chi^2$ fit of an empirical model to
the data (Fig.~\ref{Fig:WeightedDSCandidates}). The $D_s^\pm$ meson mass distribution is well modelled  by two
Gaussian functions constrained to have the same mean, but with different
 widths and  normalizations. There is negligible peaking background under the \Ds\ peak. A second peak in the $M(K^+K^-\pi^\pm)$
distribution corresponding to the Cabibbo-suppressed $D^\pm \rightarrow \phi \pi^{\pm}$ 
decay  is  also  modelled by two Gaussian functions with widths
set to those of the \Ds\ meson model scaled by the ratio of the fitted $D^\pm$ and
\Ds\ masses. The combinatoric background  is modelled by a $5^{\rm
th}$-order  polynomial function. Partially reconstructed decays such as
$\Ds \to \phi \pi^\pm \pi^0$ where the $\pi^0$ is not reconstructed are modeled
with a threshold function that extends to the \Ds\ mass after the
$\pi^0$ mass has been subtracted, given by $T(m) = \arctan\left[p_1  (m
c^2 - p_2 ) \right] + p_3$, where $p_i$ are fit parameters. In the
 fit $p_1$ is fixed to the value obtained from simulation while
the other parameters are allowed to vary.  
 
\begin{figure}[htbp]
\includegraphics[width=1.0\columnwidth]{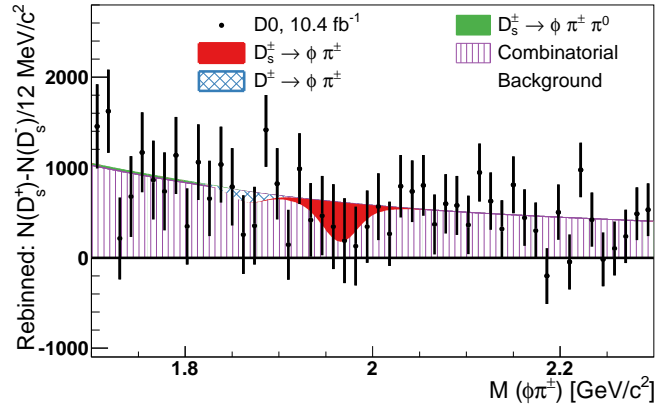}
\caption{\label{Fig:WeightedDSCandidatesDifference} 
The fit to the differences between the numbers of $\Dsp$ and $\Dsm$
mesons as function of the $\phi\pi^{\pm}$ mass (for clarity the data has
been rebinned). }
\end{figure}

The raw asymmetry  (Eq.~\ref{raw}) is extracted by fitting the
$M(K^+K^-\pi^\pm)$ distribution of the $D_s^\pm$ candidates using a
$\chi^2$ minimization. The fit is performed simultaneously, using the
same models, on the sum (Fig.~\ref{Fig:WeightedDSCandidates}) and the
difference (Fig.~\ref{Fig:WeightedDSCandidatesDifference}) of the
$M(K^+K^-\pi^+)$ distribution for the \Dsp\ candidates and the
$M(K^+K^-\pi^-)$ distribution for the \Dsm\ candidates. The functions
used to model the two distributions are
\begin{align}
W_{\rm{sum}} = & W_{D_s}  + W_D + W_{\rm{comb}} + W_{\rm{part}}, \\
W_{\rm{diff}} = &  A_{D_s} W_{D_s} + A_{D}W_{D} + A_{\rm{comb}}W_{\rm{comb}} +A_{\rm{part}}W_{\rm{part}},
\end{align}
where $W_{D_s}, W_D$, $W_{\rm{comb}}$, and $W_{\rm{part}}$ describe the
\Ds\ and $D^\pm$ mass peaks, the combinatorial background, and the
partially reconstructed events, respectively. The asymmetry of the
$D^\pm$ mass peak is $A_{D}$,  $A_\text{comb}$ is the asymmetry of the
combinatorial background, and $A_{\rm{part}}$ is the asymmetry of the
partially reconstructed events. 
 
The result of the fit is shown in Figs.~\ref{Fig:WeightedDSCandidates}
and \ref{Fig:WeightedDSCandidatesDifference} with a total $\chi^2 =
171$ for 179 degrees of freedom corresponding to a $p$-value of 0.65.
The number of signal events in the sample is $N(D_s^\pm) = 452,\!013
\pm 1,\!866$ and the fitted asymmetry parameters are $ A_{D_s} = (-0.43
\pm 0.26)\%$, $A_{D} = (-0.31 \pm 0.67)\%$,  $A_\text{comb} = (0.46 \pm
0.04)\%$, and $A_{\rm{part}} = (0.4 \pm 2.1)\%$. The  value
of the background asymmetry, $A_\text{comb}$, is consistent with 
approximately half the combinatoric background being $K^+K^-K^\pm$ or 
$K^\pm\pi^+\pi^-$ events with an average kaon reconstruction asymmetry of 1\%.

To test the sensitivity and accuracy of the fitting procedure, the sign
of the charge of the pion is randomised in the data set used in the analysis to introduce an asymmetry
signal. We simulate a range of raw signals with asymmetries from
$A_{D_s} = -2.0\%$ to $+2.0\%$ in steps of $0.2\%$, and $A_{D}$ from
$-2.0\%$ to $+2.0\%$ in steps of $0.5\%$ with 1000 pseudo-experiments
performed for each step. Each pseudo-experiment is performed with the
same statistics as the measurement. No significant systematic biases are found, and
the uncertainties are consistent with the expectation due to the sample
size.

Systematic uncertainties of the fitting method are evaluated by varying
the fitting procedure. The mass range of the fit is shifted from $1.700
<  M(K^+K^-\pi^\pm) < 2.300$~GeV$/c^2$ to $1.724 <  M(K^+K^-\pi^\pm) <
2.270$~GeV$/c^2$ in steps of 6~MeV$/c^2$   resulting in an uncertainty 
on the asymmetry of $0.044\%$. The functions modelling the signal are
modified to fit the $D^\pm$ and $D_s^\pm$ mass peaks by single Gaussian
functions, the background is fitted by varying between a $4^{\rm th}$
and $7^{\rm th}$ order polynomial function, and the parameter $p_1$ in
the threshold function is allowed to vary.  This yields an uncertainty
on the asymmetry of $0.008\%$. The width of the mass bins is changed
between 1 and 12~MeV$/c^2$ resulting in an uncertainty of $0.033\%$.
The systematic uncertainty is assigned to be half of the maximal
variation in the asymmetry for each of these sources  added in
quadrature. The total effect of these 
systematic sources of uncertainty is a systematic uncertainty of 
$0.056\%$ on the raw asymmetry $A_{D_s}$.
 
As a cross-check  variations of the various asymmetry
models are also examined. The asymmetries introduced by the functions
used to model the threshold behaviour and the combinatoric background
are set to the same value,  $A_\text{comb} = A_\text{part}$. In a separate
check the asymmetry of the threshold function is set to zero. Given the
statistical and systematic uncertainties, the observed variations of
$0.009\%$ can be neglected.

The residual detector tracking charge asymmetry has been studied in
Refs.~\cite{d0assl,d0adsl,dimuon1} using $\ks \rightarrow \pi^+\pi^-$
and $K^{\ast\pm} \rightarrow \ks \pi^\pm$ decays. After polarity weighting, 
no significant
residual track reconstruction asymmetries have been found, and no
correction for tracking asymmetries needs to be applied. The tracking
asymmetry of charged pions has been found to be less than $0.05\%$ using
MC simulations which is assigned as a systematic uncertainty.

Any asymmetry between the reconstruction of $K^+$ and $K^-$ mesons 
cancels as we require that the two kaons form a $\phi$ meson. However,
there is a small residual asymmetry in the momentum of the kaons
produced by the decay of the $\phi$ meson due to $\phi$-$f_0(980)$
interference~\cite{bellePhi}. The kaon asymmetry is measured using the
decay $K^{\ast 0} \rightarrow K^+\pi^-$~\cite{d0adsl} and is used to
determine the residual asymmetry due to this interference, 
$A_{KK} = \left[ -0.042 \pm 0.023 \thinspace (\mbox{syst})\right]\%$.

The charge asymmetry introduced by requiring the data to satisfy muon
triggers needs to be included in the overall detector asymmetry.
The effect of the  residual reconstruction asymmetry of the muon system
has been measured using $J/\psi \rightarrow \mu^+\mu^-$ decays as
described in Ref.~\cite{d0adsl}. This asymmetry is determined as a function
of $p_T^\mu$ and $|\eta^\mu|$, and the final correction is obtained by a
weighted average over the normalized ($p_T^\mu$, $|\eta^\mu|$) yields,
as determined from fits to the $M(K^+K^-\pi^\pm)$ distribution. The
resulting correction is $A_\mu = \left[ -0.036 \pm 0.010 \thinspace
(\mbox{syst})\right]\%$.

The combined residual detector asymmetry correction is
\begin{equation}
A_{\rm det} = A_\mu + A_{KK} =  \left[ -0.078 \pm 0.056 \thinspace (\mbox{syst})\right]\%,
\end{equation}
which includes the $0.05\%$ systematic uncertainty on the residual
asymmetry in track reconstruction. The remaining corrections are the
physics background asymmetries contributing to $A_{\rm phys}$, 
which are the only corrections extracted from 
MC  simulation.  The \Ds\ signal decays can also be produced in the
decay chain of 
$b$ hadrons. 
We assume that the decays of excited \Ds\ states
proceed via the strong and electromagnetic interactions and do not
introduce any CPV.

Most decays of \Bs\ mesons result in the production  of a \Ds\ meson.
These can be grouped into three categories. Semileptonic decays, $\Bs
\rightarrow \ell^+ \nu \Dsm X$, have a non-zero time-integrated
flavor-specific semileptonic charge asymmetry of $\asls = \left[ -0.79
\pm 0.43 \right]\%$ obtained by taking the average of direct and
indirect measurements~\cite{d0assl,lhcbassl,dimuon2013,hfag,d0adsl}. The
correction for this asymmetry is given by the product of the fraction of
\Ds\ events produced by semileptonic \Bs\ decays, $f_{\Bs}$, and the
fraction of \Bs\ events that have mixed, $F_{\Bs}^{\text{osc}}$. Since
\asls\ is proportional to $N_{\Dsm} - N_{\Dsp}$, it has the opposite
sign to $A_{D_s}$. The second category are  \Bs\ meson decays to a pair of
\Ds\ mesons which have no effect on the measured value of \Acp\ since equal
numbers of \Dsp\ and \Dsm\ are produced. The remaining category are
hadronic decays producing one \Ds\ meson, $\Bs \to D_s^{\pm} X$. 
Since 93\%~\cite{hfag} of \Bs\ decays produce a \Ds\ meson, any net asymmetry
 will be small. The
contributions of this process to the charge asymmetries in the production
 of \Ds\ mesons are assumed to be small and are neglected.

The remaining $b$ hadron decay processes that contribute to \Acp\ are:
$\Bd  \to D_s^{\pm} X$, $B^{\pm}  \to D_s^{\pm} X$, and the small number
of $b$ baryon and $B_c$ meson decays. It is assumed that any CPV in
these decays has a negligible effect on the measurement.

 To determine $A_{\rm phys}$, a MC sample is created using the {\sc pythia}
 event generator~\cite{pythia} modified to use {\sc evtgen}~\cite{evtgen} 
 for the decay of hadrons containing $b$ and $c$
 quarks. The {\sc pythia}  inclusive jet production model is used.
 Events recorded in random beam crossings are overlaid on the simulated
 events to emulate the effect of additional collisions in the same bunch 
 crossing.  These events are processed by the full
 simulation chain, and by the same reconstruction and selection
 algorithms as used  for  data. Events are selected that contain at
 least one $D_s^\pm \rightarrow  \phi \pi^\pm$; $\phi \rightarrow K^+
 K^-$ decay. Each event is  classified based on the decay chain that is
 matched to the reconstructed particles.

The effects of  trigger
selection and track reconstruction are estimated by weighting by the
 $p_T$ of the reconstructed \Ds\ to match the distribution of the
data. 
The trigger and offline streaming requirements  are accounted for by
requiring a reconstructed muon in each of the MC events and weighting
the muons to match the $p_T^\mu$-$\eta^\mu$ distributions of muons in
the data. The weights are  obtained by taking the ratio of the muon 
$p_T^\mu$-$\eta^\mu$ distributions in the selected data sample and a
sample obtained using the zero-bias trigger condition. These weights are
then applied to the MC simulation. 
 
A large fraction of the data were collected at high instantaneous
luminosities, and there is some probability that the muon and
the \Ds\ candidate originate from different  proton-antiproton
collisions. This probability is determined by measuring the separation
along the $z$ axis of the intersection of the \Ds\ trajectory and the
track of the highest $p_T$ muon in the event. The fraction of events
that come from separate $p\bar{p}$ interactions is estimated to be 
$6.4\%$. This effect is accounted for in the simulation. 

From these studies, the sample is predicted to be composed 
of 71\% \Ds\ mesons produced directly, 
10\% from the hadronic decays of \Bs mesons (which includes $\Bs \to D_s^{\pm(\ast)}D_s^{\mp(\ast)}$),
6\% each from  the decay of $B^\pm$ and \Bd\ mesons,
and 1\%  from the decay of $b$ baryons. 
The  fraction of events that originate from \Bs\ semi-leptonic decays is
found to be $f_{\Bs} = 5.8\%$ and the fraction that have oscillated to
be $ F_{\Bs}^{\text{osc}} = 50\%$. In addition to the MC statistical
uncertainty, the systematic uncertainty on $A_{\rm{phys}}$ is determined
by varying the following quantities by  their uncertainties: the
branching ratios and production fractions of $B$ and $D$ mesons, the $D$
and $B$-meson lifetimes, and $\Delta\Gamma_s$. The largest sources of
uncertainty are the fraction of events in which a $c$ quark forms a \Ds\
meson, $f(c\to\Ds) = 0.080 \pm 0.017$~\cite{pdg2008}, and the
semileptonic branching fraction of \Bs\ mesons, ${\cal B}(\Bs
\rightarrow \ell^+ \nu \Dsm X) = (9.5 \pm 2.7)\%$. The uncertainty on
the correction due to \asls\ is $0.024\%$, yielding an asymmetry
resulting from the decay of $b$ hadrons into \Ds\ mesons of:
\begin{equation}
A_{\rm phys} = \left[ 0.023 \pm 0.024 \thinspace (\mbox{syst})\right]\%.
\end{equation}

Several consistency checks are performed by dividing the data into smaller samples 
using additional selections based on data-taking periods, magnet polarities, 
\Ds\ transverse momentum, and \Ds\ pseudo-rapidity. 
The resulting variations of \Acp\  are statistically consistent with 
the results of Eq.~\ref{eq:finalResult} (see below).

The selection criteria applied in this analysis preferentially select
the P-wave decay, \Dsdecay , over the continuum process $\Ds \to
K^+K^-\pi^\pm$ and other processes that result in a $K^+K^-\pi^\pm$
final state. In particular the helicity angle between the $D_s^\pm$ and
$K^{\mp}$ momenta in the center-of-mass frame of the $\phi$ meson and
the invariant mass $M(K^+ K^-)$ used in the likelihood ratio select
\Dsdecay\ decays. To study the possible effect of other non-P-wave 
contributions, these variables are removed
from the likelihood ratio and  replaced with the requirement $1.01 <
M(K^+ K^-) < 1.03$~GeV/$c^2$. The analysis is reoptimised and the
asymmetry is found to be $\left[ -0.63 \pm 0.35 \thinspace
(\text{stat}) \pm 0.08 \thinspace (\text{syst}) \right]\%$ which is
consistent with the main analysis and with the SM expectation of zero CPV.

The systematic uncertainty due to the fitting procedure (0.056\%) added in quadrature with
the uncertainties on $A_{\rm det}$ (0.056\%) and $A_{\rm phys}$ (0.024\%) results in a total
systematic uncertainty of $0.08\%$. The  direct {\it CP}-violating
charge asymmetry in \Ds\  mesons is found to be
\begin{equation}
\label{eq:finalResult}
\Acp = \left[ -0.38 \pm 0.26 \thinspace (\text{stat}) \pm 0.08 \thinspace (\text{syst}) \right]\% ,
\end{equation}
corresponding to a total absolute uncertainty of $0.27\%$. This is the most
precise measurement of direct CPV in the decay ${D_s^\pm \to \phi
\pi^\pm}$, and the result is in agreement with the SM expectation of zero
CPV in this decay.

%
We thank the staffs at Fermilab and collaborating institutions,
and acknowledge support from the
DOE and NSF (USA);
CEA and CNRS/IN2P3 (France);
MON, NRC KI and RFBR (Russia);
CNPq, FAPERJ, FAPESP and FUNDUNESP (Brazil);
DAE and DST (India);
Colciencias (Colombia);
CONACyT (Mexico);
NRF (Korea);
FOM (The Netherlands);
STFC and the Royal Society (United Kingdom);
MSMT and GACR (Czech Republic);
BMBF and DFG (Germany);
SFI (Ireland);
The Swedish Research Council (Sweden);
and
CAS and CNSF (China).


\begin{thebibliography}{99}
\bibitem{eta}  
The pseudorapidity is given by
$\eta=-\ln[\tan(\theta/2)]$,  $\theta$ is
the polar angle between the track momentum and the proton beam direction, and $\phi$ is the azimuthal angle of the track.
\end{thebibliography}
\end{document}